\documentclass[
preprint
]{revtex4}
\usepackage{graphicx}
\usepackage{dcolumn}
\usepackage{bm}
\usepackage{amsmath}
\usepackage{float}

\begin{document}

\title{Length Dependent Thermal Conductivity Measurements Yield Phonon Mean Free Path Spectra in Nanostructures}

\pacs{ }


\author{Hang Zhang}
\affiliation{Division of Engineering and Applied Science\\
California Institute of Technology\\
Pasadena, CA 91125}

\author{Chengyun Hua}
\affiliation{Division of Engineering and Applied Science\\
California Institute of Technology\\
Pasadena, CA 91125}

\author{Ding Ding}
\affiliation{Division of Engineering and Applied Science\\
California Institute of Technology\\
Pasadena, CA 91125}

\author{Austin J. Minnich}
\email{aminnich@caltech.edu}
\affiliation{Division of Engineering and Applied Science\\
California Institute of Technology\\
Pasadena, CA 91125}


\begin{abstract}

Thermal conductivity measurements over variable lengths on nanostructures such as nanowires provide important information about the mean free paths (MFPs) of the phonons responsible for heat conduction. However, nearly all of these measurements have been interpreted using an average MFP even though phonons in many crystals possess a broad MFP spectrum. Here, we present a reconstruction method to obtain MFP spectra of nanostructures from variable-length thermal conductivity measurements. Using this method, we investigate recently reported length-dependent thermal conductivity measurements on SiGe alloy nanowires and suspended graphene ribbons. We find that the recent measurements on graphene imply that 70\% of the heat in graphene is carried by phonons with MFPs longer than 1 micron.
 
\end{abstract}

\maketitle

Thermal transport in nanostructures has been a topic of intense interest in recent years\cite{Zebarjadi2012,Cahill2014, Zhao2014}. When the characteristic dimensions of nanostructures such as the diameter of a nanowire approach phonon mean free paths (MFPs), the thermal conductivity can be substantially smaller than the bulk value due to scattering from sample boundaries. Significant thermal conductivity reductions have been observed in a number of nanoscale systems, including nanowires\cite{ Li2003, Hochbaum2008, Boukai2008}, nanotubes\cite{Chang2008}, thin Si membranes\cite{Johnson2013}, and micron size beams at cryogenic temperatures\cite{ Tighe1997}. This concept has been widely adopted in thermoelectrics applications\cite{ Poudel2008,Biswas2012, Venkatasubramanian2009,Tian2013}.\\

Understanding and engineering the thermal conductivity reduction in nanostructures requires knowledge of phonon scattering mechanisms in the form of the phonon MFPs. The MFP accumulation function, which we term the MFP spectrum in this work, has been demonstrated to be a particularly useful quantity to describe the values of the MFPs relevant for heat conduction\cite{DamesCRC}. In several works, information about MFPs was obtained by measuring the thermal conductivity over variable lengths of nanostructures such as nanotubes\cite{ Chang2008}, graphene ribbons\cite{ Xu2014} and SiGe nanowires\cite{ Hsiao2013}. If phonons have MFPs exceeding the distance between the heat source and sink their contribution to thermal conductivity is reduced compared to that in the bulk material, and thus the deviations of the measured thermal conductivity from the bulk value provide information on the phonon MFPs. However, prior studies extracted only an average MFP despite the fact that recent works have demonstrated that in many solids phonon MFPs vary over orders of magnitude, making the approximation of an average MFP for all phonons quite poor\cite{Esfarjani2011,Henry2008}. 

In principle, information about the full MFP spectrum should be contained in these variable-length thermal conductivity measurements, just as the MFP spectrum can be obtained from thermal conductivity measurements performed over variable thermal length scales in MFP spectroscopy\cite{Minnich2011PRL}.  In particular, the method proposed by Minnich based on convex optimization\cite{ Minnich2012} should be applicable to the present situation provided that the suppression function that describes the effect of the finite length on the thermal conductivity can be identified. Li et al obtained the phonon MFP spectrum of graphite along the c-axis from thickness dependent thermal conductivities obtained with molecular dynamics simulations\cite{Wei2014}, but their suppression function was not rigorously obtained from the Boltzmann Transport Equation (BTE). 

In this report, we present a reconstruction approach to obtain MFP spectra from variable-length thermal conductivity measurements. We use a recently reported analytical solution of the BTE, along with efficient numerical simulations, to identify a suppression function that describes the discrepancy between the actual heat flux and that predicted by Fourier's law for a finite length domain. The MFP spectrum is then obtained using the convex optimization method described in Ref. \citenum{Minnich2012}. We apply this approach to SiGe nanowires and graphene ribbons. The measurements on graphene ribbons imply that MFPs are exceedingly long, with 70\% of the heat being carried by phonons with MFPs longer than 1 micron.

\section*{Theory}

Our goal is to relate experimentally measured thermal conductivities to the MFP spectrum, or the accumulated thermal conductivity as a function of MFP\cite{DamesCRC}. Following the approach of Ref. \citenum{Minnich2012}, we therefore seek an equation of the form:\\
\begin{equation}
k=\int_{0}^{\infty}S(\text{Kn}_{\omega})f(\Lambda_\omega) d\Lambda_{\omega}=\int_{0}^{\infty} L^{-1} K(\text{Kn}_{\omega})F(\Lambda_\omega) d\Lambda_{\omega}
\end{equation}
where $\text{Kn}_{\omega}=\Lambda_{\omega}/L$ is the Knudsen Number, $\Lambda_\omega$ is the MFP, L is the sample length along the direction of temperature gradient, $k$ denotes thermal conductivity as a function of length $L$, $f(\Lambda_\omega)$ and $F(\Lambda_\omega)$ are differential and accumulative MFP spectra related by $F(\Lambda_\omega)=\int_{0}^{\Lambda_\omega}f(\Lambda) d\Lambda$, and $S$ is the heat flux suppression function that equals the ratio of actual heat flux to the Fourier's law prediction. The kernel $K$ is defined as $K(\text{Kn}_{\omega})=-dS/d\text{Kn}_{\omega}$.

The inputs to this equation are a finite number of measured thermal conductivities $k$ as a function of lengths $L$. To close the problem, we must identify the suppression function by solving the BTE, given by\cite{Majumdar1993}:\\
\begin{equation}
\frac{\partial e_{\omega}}{\partial t}+\textbf{v}\cdot\bigtriangledown_{\textbf{r}}e_{\omega}=-\frac{e_{\omega}-e^{0}_{\omega}}{\tau_{\omega}}
\end{equation}
where $e_{\omega}$ is the desired distribution function, $\omega$ is the angular frequency, $e^{0}_{\omega}$ is the distribution function at the equilibrium state, $\textbf{v}$ is the group velocity of phonons, and $\tau_{\omega}$ is the relaxation time of phonons at certain frequency.\\

We obtain this function using two distinct approaches: a semi-analytic method and a numerical Monte Carlo (MC) method. First, we use a recently reported semi-analytical solution for steady heat conduction through a crystal of thickness $L$ with two blackbody boundaries\cite{ Hua2014PRB}. In this solution, the BTE is linearized and solved using a series expansion method. The full details are given in Ref. \citenum{ Hua2014a}. The final result for the suppression function and kernel are:
\begin{equation}
S (\text{Kn}_{\omega}) = 1 + 3\text{Kn}_{\omega}\left[E_5(\text{Kn}_{\omega}^{-1})-\frac{1}{4}\right]
\end{equation}
\begin{equation}
K(\text{Kn}_{\omega}) = - \frac{dS }{d\text{Kn}_{\omega}} = \frac{3}{4}-3 E_5(\text{Kn}_{\omega}^{-1})-\frac{3}{\text{Kn}_{\omega}}E_4(\text{Kn}_{\omega}^{-1})
\end{equation}\\
 Where, $E_n(x)$ is the exponential integral function, given by: $E_n(x)=\int^{1}_{0}\mu^{n-2}exp(-\frac{x}{\mu})d\mu$\cite{Chen2005book}.

This equation was derived by neglecting temperature slip between the black walls. Physically, this assumption is similar to the weakly quasiballistic regime described in Ref. \citenum{Hua2014PRB} and implies that the ballistic phonons are low frequency modes with a small heat capacity. The assumption has been shown to be quite accurate for experimentally accessible length scales\cite{ Hua2014a}.

We plot this result in Fig.1a. For extremely short MFPs compared to the sample length $L$, the suppression function equals unity, indicating these phonons are diffusive and their heat flux contribution equals the Fourier's law prediction. As the Knudsen number increases, the suppression function decreases and eventually approaches zero, indicating that phonons contribute a smaller amount to the heat flux than predicted by Fourier's law. Physically, this suppression occurs because phonons cannot travel a full MFP before being absorbed by the blackbody boundary.

We additionally solve the BTE numerically to validate the calculations above as well as to consider more complex situations such as when boundary scattering occurs. For this calculation, we use a linearized deviational Monte Carlo method to solve the adjoint BTE as described by P\'eraud et al\cite{Peraud2011}. This technique solves BTE by simulating advection and scattering of particles that represent phonons traveling inside the simulation domain. Substantial reductions in computational cost are achieved through a number of simple changes to the original MC algorithm. First, the deviational algorithm simulates only the deviation from a known equilibrium Bose-Einstein distribution, thereby incorporating deterministic information and reducing the variance. Further, for small temperature differences, the collision term in the BTE can be linearized, allowing particles to be simulated completely independently and without spatial and temporal discretization\cite{Peraud2012}. Next, we use a variable local equilibrium temperature method that closely matches the steady-state temperature profile.  

Finally, we solve the adjoint BTE rather than the traditional BTE\cite{Peraud2014,Peraud2014a}. In the original algorithms of Ref. \citenum{Peraud2011,Peraud2012}, the probability for a certain phonon mode to be sampled is proportional to the density of states. Therefore, low frequency phonons are rarely sampled even though they contribute substantially to thermal conductivity, leading to large stochastic noise. The adjoint method overcomes this limitation by drawing particles with equal probability among all phonon modes and correcting the bias introduced by this sampling when thermal properties are calculated. With these advances in numerical approach, we are able to solve the BTE in a 100 micron long domain in minutes on a desktop computer. Further, this numerical approach can incorporate boundary scattering mechanisms for arbitrary geometries exactly, unlike the analytical treatment.\\

To validate the code, we calculate the MFP spectrum for an infinite planar slab with two blackbody boundary conditions, the same problem solved by the semi-analytical method. For this calculation, we use an isotropically averaged dispersion for Si. The original dispersion and relaxation times were calculated by density functional theory (DFT) by Jes\'us Carrete and N. Mingo with ShengBTE\cite{Li2014,shengbte} and Phonopy\cite{phonopy}, from interatomic force constants obtained with VASP\cite{ Kresse1993, Kresse1994, Kresse1996, Kresse1996a}. We reduce computational cost by taking advantage of the cubic symmetry of Si and computing an isotropic equivalent dispersion as described in Ref. \citenum{Hua2014PRB2}. Using this dispersion, we calculate the MFP spectrum for variable lengths, as in Fig 1b. We observe that decreasing the length of the domain results in the suppression of long MFP phonons to thermal conductivity compared to the bulk spectrum. The ratio of the differential MFP spectrum for a finite length to that for an infinite length yields the suppression function and is plotted for two lengths in Fig. 1a, demonstrating that the function obtained from our numerical approach exactly agrees with the analytical result.

With these tools, we now demonstrate the principal result of this work by using the suppression function to reconstruct the MFP spectrum from variable-length thermal conductivity measurements on a Si slab. We synthesized thermal conductivities as a function of length as shown in Fig. 1c. With these length dependent thermal conductivities and the suppression function, we used the same convex optimization method introduced in Ref. \citenum{Minnich2012} to reconstruct the MFP spectrum. As in Figure 1d, the reconstructed accumulative MFP distribution is in excellent agreement with the actual accumulative MFP distribution, which is obtained from the DFT calculation, demonstrating that our approach can accurately reconstruct the MFP spectrum from length-dependent thermal conductivities of Si slabs.

We now numerically demonstrate that our approach can be applied to more general problems than Si slabs with a single scattering mechanism. We consider a SiGe nanowire in which point defects and the nanowire boundaries scatter phonons in addition to the intrinsic phonon-phonon scattering mechanism. The phonon-phonon relaxation times are taken to be the same as those of pure Si\cite{Li2014,shengbte,phonopy,Kresse1993, Kresse1994, Kresse1996, Kresse1996a}, while the mass defect scattering rate is given by $\tau_{\omega,i}^{-1}=x(1-x)A\omega^4$\cite{ Wang2010}, where $A$ is a constant of $3.01\times 10^{-41}$ $\text s^3$ for Si$_{1-x}$Ge$_{x}$, which is obtained from Ref. \citenum{ Wang2010}. This model predicts a thermal conductivity of 14 W/mK for bulk Si$_{0.9}$Ge$_{0.1}$, which is consistent with other models by DFT calculation and experimental result\cite{ Garg2011,Wang2010}. These scattering rates are combined using Matthiessen's rule. We incorporate boundary scattering by explicitly simulating phonon trajectories inside a nanowire with a square cross-section of size 100 nm by 100 nm. We use Ziman's specularity parameter, $ p=\exp(-16\pi^2\sigma^2/\lambda^2)$, to determine the probability of specular or diffuse scattering, where $\sigma$ is surface roughness and $\lambda$ is the phonon wavelength\cite{Ziman1962}.

With this framework, we use our MC simulations to calculate the length-dependent thermal conductivities for a Si$_{0.9}$Ge$_{0.1}$ nanowire with surface roughness of $\sigma$ = 0.1 nm, a 100 nm by 100 nm square cross-section, over lengths from $L$ = 5 nm to 16 mm as would be obtained in an experiment. Then, using only our knowledge of these thermal conductivities and the suppression function, we perform the reconstruction procedure to obtain the MFP spectrum of the nanowire. This result is shown in Fig 2a. The reconstructed spectrum is in good agreement with the actual one without requiring any knowledge of the scattering mechanisms in the nanowire. Thus, the success of the reconstruction of MFP spectrum of the nanowire demonstrates the self-consistency of our approach.

\section*{Discussion}

We now use our approach to examine two recent reports of length-dependent thermal conductivities in nanostructures. First, we consider SiGe alloy nanowires as investigated by Hsiao et al\cite{ Hsiao2013}. These nanowires were reported to have ballistic heat conduction persisting over approximately 8 microns. To investigate this experimental report, we calculate the length-dependent thermal conductivities of Si$_{0.9}$Ge$_{0.1}$ nanowires, which has the approximately the same cross-sectional area, 100 nm by 100 nm cross section, as that of the nanowires in Ref. \citenum{Hsiao2013}, whose diameters range from 50 nm to 180 nm. Simulated data are shown in Fig 2b for nanowires with extremely rough boundaries, $\sigma \rightarrow \infty$, and smooth boundaries, with $\sigma$=0.1 nm.  

We observe a discrepancy between the trends of experimental data and our simulations. The experimental data suggest that the thermal conductivities of these SiGe nanowires are mainly due to phonons with a narrow MFP spectrum around $\sim$ 8.3 $\mu$m, but our simulations indicate that even for very smooth nanowires ($\sigma$ = 0.1 nm), phonons within this range only contribute $\sim$ 15\% of the total thermal conductivity (Fig. 2a). Most of the heat is carried by MFPs less than 1 micron, with some heat being carried by longer MFPs in the partially specular case. Additionally, most actual nanowires have surface roughness higher than 0.1 nm\cite{Azeredo2013,Feser2012}, and in this case long MFP phonons contribute only a small amount to heat conduction. Due to the relatively large diameter of the nanowire, changes to the phonon dispersion for thermal phonons due to phonon confinement are unlikely\cite{ Turney2010}. The experimental measurements thus do not agree with our self-consistent calculations. Due to the lack of appropriate experimental reports, we are unable to apply our approach to other nanowire data sets.  Further experimental investigation is necessary to address this discrepancy.

Next, we consider recent measurements on graphene\cite{ Xu2014}. In this work, Xu et al\cite{Xu2014} performed thermal conductivity measurements over variable lengths on suspended single-layer graphene ribbons to infer an average MFP of 240 nm at room temperature. Using our approach, we can use these same measurements to obtain the MFP spectrum of graphene. Due to the difficulty of fabrication for very long suspended graphene devices and vulnerability of these devices during measurement, the authors of Ref. \citenum{Xu2014} didn't obtain the saturated thermal conductivity from an extremely long, or ``bulk'', graphene sample. Therefore, we evaluated the saturation value as $\sim$ 2,000 W/mK with the same extrapolating method in Ref. \citenum{Wei2014}, which is in the range of previous reported experimental results\cite{Cai2010,Chen2011,Chen2012}. 

There are two subtleties that require discussion before applying our reconstruction approach to graphene. First, our derivation is based on the relaxation time approximation (RTA) of the BTE. It is well known that the RTA with the computed relaxation times from DFT underpredicts the thermal conductivity of graphene due to the importance of normal processes\cite{Lindsay2010}. However, while our derivation is based on the RTA, we do not make any assumption of the values of the relaxation times, but rather only that an effective relaxation time for each phonon mode can be identified. Our approach can be applied to graphene provided that we regard the MFP variable as an effective MFP that represents the average propagation length for a particular phonon frequency as determined by both normal and Umklapp processes. 

Second, we have derived our suppression function for an isotropic material with a three-dimensional phase space. While graphene can be reasonably modeled as isotropic for the in-plane directions, the phase space is two-dimensional. This dimensionality change requires a modification of the form of the BTE for 2D materials. Repeating the derivation in Ref. \citenum{ Hua2014a} for a 2D phase space yields the suppression function $S$ and kernel $K$ as:
\begin{equation}
S_{2D}(\text{Kn}_{\omega}) = 1 + \frac{4}{\pi}\text{Kn}_{\omega}\left[\int_0^{\frac{\pi}{2}} \text{cos}^3(\theta) e^{-\frac{1}{\text{Kn}_{\omega}\text{cos}(\theta)}}d\theta-\frac{2}{3}\right]
\end{equation}
\begin{equation}
K_{2D}(\text{Kn}_{\omega}) = - \frac{dS_{2D}}{d\text{Kn}_{\omega}} = \frac{8}{3\pi}-\frac{4}{\pi}\int_0^{\frac{\pi}{2}} \text{cos}^3(\theta) e^{-\frac{1}{\text{Kn}_{\omega}\text{cos}(\theta)}}d\theta-\frac{4}{\pi\text{Kn}_{\omega}}\int_0^{\frac{\pi}{2}} \text{cos}^2(\theta) e^{-\frac{1}{\text{Kn}_{\omega}\text{cos}(\theta)}}d\theta
\end{equation}\\

Figure 3a plots the 2D and 3D suppression functions and kernels, demonstrating that the two are quite similar. Using the 2D kernel, we apply our method to obtain the MFP spectrum of a graphene ribbon as in Fig. 3b. This result shows that MFPs in graphene span a large range from $\sim$ 100 nm to $\sim$ 10 $\mu$m . In addition, we observe that a large portion of heat in graphene is carried by long MFP phonons: $\sim$ 70\% of thermal conductivity are from phonons with MFPs greater than 1 $\mu$m. Further, as reported in Ref. \citenum{Xu2014}, the widths of these graphene ribbons are only between 2 to 4 microns. From Fig. 3b, phonons with MFPs longer than 4 $\mu$m still carry 13\% of heat, which means that some of these long MFP phonons must be specularly reflected at the edges of the graphene ribbons. This observation further confirms the report in Ref. \citenum{Xu2014} of weak width-dependent thermal conductivities of suspended graphene ribbons when widths are larger than 1.5 $\mu$m. Our MFP reconstruction approach has thus provided valuable insights into the intrinsic and edge scattering mechanisms in graphene ribbons that are difficult to obtain from knowledge of only the average MFP.
 
\section*{Summary}
We have presented a reconstruction method that allows MFP spectra of nanostructures to be obtained from length-dependent thermal conductivity measurements. Our approach requires no prior knowledge of the scattering mechanisms in the nanostructure. By applying our approach to recent measurements on graphene ribbons, we find that more than half of the heat in graphene is contributed by phonons with MFPs exceeding 1 micron.
\\

\section*{Acknowledgments}
This work was supported by a start-up fund from the California Institute of Technology and by the National Science Foundation under CAREER Grant CBET 1254213.\\

\section*{Author contributions}
H.Z., C.H. and D.D. performed simulations and calculations. H.Z. and A.M. analyzed data. H.Z., A.M.,C.H. and D.D. discussed the result. H.Z. and A.M. wrote the main manuscript text. 

\section*{Additional Information}
Competing financial interests: The authors declare no competing financial interests

\bibliographystyle{naturemag}


\begin{thebibliography}{10}
\expandafter\ifx\csname url\endcsname\relax
  \def\url#1{\texttt{#1}}\fi
\expandafter\ifx\csname urlprefix\endcsname\relax\def\urlprefix{URL }\fi
\providecommand{\bibinfo}[2]{#2}
\providecommand{\eprint}[2][]{\url{#2}}

\bibitem{Zebarjadi2012}
\bibinfo{author}{Zebarjadi, M.}, \bibinfo{author}{Esfarjani, K.},
  \bibinfo{author}{Dresselhaus, M.~S.}, \bibinfo{author}{Ren, Z.~F.} \&
  \bibinfo{author}{Chen, G.}
\newblock \bibinfo{title}{Perspectives on thermoelectrics: from fundamentals to
  device applications}.
\newblock \emph{\bibinfo{journal}{Energy Environ. Sci.}}
  \textbf{\bibinfo{volume}{5}}, \bibinfo{pages}{5147--5162}
  (\bibinfo{year}{2012}).

\bibitem{Cahill2014}
\bibinfo{author}{Cahill, D.~G.} \emph{et~al.}
\newblock \bibinfo{title}{Nanoscale thermal transport. ii. 2003-2012}.
\newblock \emph{\bibinfo{journal}{Appl. Phys. Rev.}}
  \textbf{\bibinfo{volume}{1}}, \bibinfo{pages}{011305} (\bibinfo{year}{2014}).

\bibitem{Zhao2014}
\bibinfo{author}{Zhao, L.-D.}, \bibinfo{author}{Dravid, V.~P.} \&
  \bibinfo{author}{Kanatzidis, M.~G.}
\newblock \bibinfo{title}{The panoscopic approach to high performance
  thermoelectrics}.
\newblock \emph{\bibinfo{journal}{Energy Environ. Sci.}}
  \textbf{\bibinfo{volume}{7}}, \bibinfo{pages}{251--268}
  (\bibinfo{year}{2014}).

\bibitem{Li2003}
\bibinfo{author}{Li, D.} \emph{et~al.}
\newblock \bibinfo{title}{Thermal conductivity of individual silicon
  nanowires}.
\newblock \emph{\bibinfo{journal}{Appl. Phys. Lett.}}
  \textbf{\bibinfo{volume}{83}}, \bibinfo{pages}{2934--2936}
  (\bibinfo{year}{2003}).

\bibitem{Hochbaum2008}
\bibinfo{author}{Hochbaum, A.~I.} \emph{et~al.}
\newblock \bibinfo{title}{Enhanced thermoelectric performance of rough silicon
  nanowires}.
\newblock \emph{\bibinfo{journal}{Nature}} \textbf{\bibinfo{volume}{451}},
  \bibinfo{pages}{163--167} (\bibinfo{year}{2008}).

\bibitem{Boukai2008}
\bibinfo{author}{Boukai, A.~I.} \emph{et~al.}
\newblock \bibinfo{title}{Silicon nanowires as efficient thermoelectric
  materials}.
\newblock \emph{\bibinfo{journal}{Nature}} \textbf{\bibinfo{volume}{451}},
  \bibinfo{pages}{168--171} (\bibinfo{year}{2008}).

\bibitem{Chang2008}
\bibinfo{author}{Chang, C.~W.}, \bibinfo{author}{Okawa, D.},
  \bibinfo{author}{Garcia, H.}, \bibinfo{author}{Majumdar, A.} \&
  \bibinfo{author}{Zettl, A.}
\newblock \bibinfo{title}{Breakdown of fourier's law in nanotube thermal
  conductors}.
\newblock \emph{\bibinfo{journal}{Phys. Rev. Lett.}}
  \textbf{\bibinfo{volume}{101}}, \bibinfo{pages}{075903}
  (\bibinfo{year}{2008}).

\bibitem{Johnson2013}
\bibinfo{author}{Johnson, J.~A.} \emph{et~al.}
\newblock \bibinfo{title}{Direct measurement of room-temperature nondiffusive
  thermal transport over micron distances in a silicon membrane}.
\newblock \emph{\bibinfo{journal}{Phys. Rev. Lett.}}
  \textbf{\bibinfo{volume}{110}}, \bibinfo{pages}{025901}
  (\bibinfo{year}{2013}).

\bibitem{Tighe1997}
\bibinfo{author}{Tighe, T.~S.}, \bibinfo{author}{Worlock, J.~M.} \&
  \bibinfo{author}{Roukes, M.~L.}
\newblock \bibinfo{title}{Direct thermal conductance measurements on suspended
  monocrystalline nanostructures}.
\newblock \emph{\bibinfo{journal}{Appl. Phys. Lett.}}
  \textbf{\bibinfo{volume}{70}}, \bibinfo{pages}{2687--2689}
  (\bibinfo{year}{1997}).

\bibitem{Poudel2008}
\bibinfo{author}{Poudel, B.} \emph{et~al.}
\newblock \bibinfo{title}{High-thermoelectric performance of nanostructured
  bismuth antimony telluride bulk alloys}.
\newblock \emph{\bibinfo{journal}{Science}} \textbf{\bibinfo{volume}{320}},
  \bibinfo{pages}{634--638} (\bibinfo{year}{2008}).

\bibitem{Biswas2012}
\bibinfo{author}{Biswas, K.} \emph{et~al.}
\newblock \bibinfo{title}{High-performance bulk thermoelectrics with all-scale
  hierarchical architectures}.
\newblock \emph{\bibinfo{journal}{Nature}} \textbf{\bibinfo{volume}{489}},
  \bibinfo{pages}{414--418} (\bibinfo{year}{2012}).

\bibitem{Venkatasubramanian2009}
\bibinfo{author}{Chowdhury, I.} \emph{et~al.}
\newblock \bibinfo{title}{On-chip cooling by superlattice-based thin-film
  thermoelectrics}.
\newblock \emph{\bibinfo{journal}{Nat. Nanotechnol.}}
  \textbf{\bibinfo{volume}{4}}, \bibinfo{pages}{235--238}
  (\bibinfo{year}{2009}).

\bibitem{Tian2013}
\bibinfo{author}{Tian, Z.}, \bibinfo{author}{Lee, S.} \& \bibinfo{author}{Chen,
  G.}
\newblock \bibinfo{title}{Heat transfer in thermoelectric materials and
  devices}.
\newblock \emph{\bibinfo{journal}{J. Heat Transfer}}
  \textbf{\bibinfo{volume}{135}}, \bibinfo{pages}{061605--061605}
  (\bibinfo{year}{2013}).

\bibitem{DamesCRC}
\bibinfo{author}{Chen, G.} \& \bibinfo{author}{Dames, C.}
\newblock \bibinfo{title}{Thermal conductivity of nanostructured thermoelectric
  materials}.
\newblock In \emph{\bibinfo{booktitle}{Thermoelectrics Handbook: Macro to
  Nano}}, \bibinfo{pages}{42--1--42--16--} (\bibinfo{publisher}{CRC Press},
  \bibinfo{year}{2005}).

\bibitem{Xu2014}
\bibinfo{author}{Xu, X.} \emph{et~al.}
\newblock \bibinfo{title}{Length-dependent thermal conductivity in suspended
  single-layer graphene}.
\newblock \emph{\bibinfo{journal}{Nat. Commun.}} \textbf{\bibinfo{volume}{5}},
  \bibinfo{pages}{1--6} (\bibinfo{year}{2014}).

\bibitem{Hsiao2013}
\bibinfo{author}{Hsiao, T.-K.} \emph{et~al.}
\newblock \bibinfo{title}{Observation of room-temperature ballistic thermal
  conduction persisting over 8.3 $\mu$m in sige nanowires}.
\newblock \emph{\bibinfo{journal}{Nat. Nanotechnol.}} \textbf{\bibinfo{volume}{8}},
  \bibinfo{pages}{534--538} (\bibinfo{year}{2013}).

\bibitem{Esfarjani2011}
\bibinfo{author}{Esfarjani, K.}, \bibinfo{author}{Chen, G.} \&
  \bibinfo{author}{Stokes, H.~T.}
\newblock \bibinfo{title}{Heat transport in silicon from first-principles
  calculations}.
\newblock \emph{\bibinfo{journal}{Phys. Rev. B}} \textbf{\bibinfo{volume}{84}},
  \bibinfo{pages}{085204} (\bibinfo{year}{2011}).

\bibitem{Henry2008}
\bibinfo{author}{Henry, A.~S.} \& \bibinfo{author}{Chen, G.}
\newblock \bibinfo{title}{Spectral phonon transport properties of silicon based
  on molecular dynamics simulations and lattice dynamics}.
\newblock \emph{\bibinfo{journal}{J. Comput. Theor. Nanos.}} \textbf{\bibinfo{volume}{5}}, \bibinfo{pages}{141--152}
  (\bibinfo{year}{2008}).

\bibitem{Minnich2011PRL}
\bibinfo{author}{Minnich, A.~J.} \emph{et~al.}
\newblock \bibinfo{title}{Thermal conductivity spectroscopy technique to
  measure phonon mean free paths}.
\newblock \emph{\bibinfo{journal}{Phys. Rev. Lett.}}
  \textbf{\bibinfo{volume}{107}}, \bibinfo{pages}{095901}
  (\bibinfo{year}{2011}).

\bibitem{Minnich2012}
\bibinfo{author}{Minnich, A.~J.}
\newblock \bibinfo{title}{Determining phonon mean free paths from observations
  of quasiballistic thermal transport}.
\newblock \emph{\bibinfo{journal}{Phys. Rev. Lett.}}
  \textbf{\bibinfo{volume}{109}}, \bibinfo{pages}{205901}
  (\bibinfo{year}{2012}).

\bibitem{Wei2014}
\bibinfo{author}{Wei, Z.} \emph{et~al.}
\newblock \bibinfo{title}{Phonon mean free path of graphite along the c-axis}.
\newblock \emph{\bibinfo{journal}{Appl. Phys. Lett.}}
  \textbf{\bibinfo{volume}{104}}, \bibinfo{pages}{081903}
  (\bibinfo{year}{2014}).

\bibitem{Majumdar1993}
\bibinfo{author}{Majumdar, A.}
\newblock \bibinfo{title}{Microscale heat conduction in dielectric thin films}.
\newblock \emph{\bibinfo{journal}{J. Heat Transfer}}
  \textbf{\bibinfo{volume}{115}}, \bibinfo{pages}{7--16}
  (\bibinfo{year}{1993}).

\bibitem{Hua2014PRB}
\bibinfo{author}{Hua, C.} \& \bibinfo{author}{Minnich, A.~J.}
\newblock \bibinfo{title}{Transport regimes in quasiballistic heat conduction}.
\newblock \emph{\bibinfo{journal}{Phys. Rev. B}} \textbf{\bibinfo{volume}{89}},
  \bibinfo{pages}{094302} (\bibinfo{year}{2014}).

\bibitem{Hua2014a}
\bibinfo{author}{{Hua}, C.} \& \bibinfo{author}{{Minnich}, A.~J.}
\newblock \bibinfo{title}{{Cross-plane heat conduction in thin solid films}}.
\newblock \emph{\bibinfo{journal}{ArXiv e-prints}}  (\bibinfo{year}{2014}).
\newblock \bibinfo{note}{Http://arxiv.org/abs/1410.2845}, \eprint{1410.2845}.

\bibitem{Chen2005book}
\bibinfo{author}{Chen, G.}
\newblock \emph{\bibinfo{title}{Nanoscale energy transport and conversion : a
  parallel treatment of electrons, molecules, phonons, and photons}}
  (\bibinfo{publisher}{Oxford University Press}, \bibinfo{address}{Oxford ; New
  York}, \bibinfo{year}{2005}).

\bibitem{Peraud2011}
\bibinfo{author}{Peraud, J.-P.~M.} \& \bibinfo{author}{Hadjiconstantinou,
  N.~G.}
\newblock \bibinfo{title}{Efficient simulation of multidimensional phonon
  transport using energy-based variance-reduced monte carlo formulations}.
\newblock \emph{\bibinfo{journal}{Phys. Rev. B}} \textbf{\bibinfo{volume}{84}},
  \bibinfo{pages}{205331} (\bibinfo{year}{2011}).

\bibitem{Peraud2012}
\bibinfo{author}{Peraud, J.-P.~M.} \& \bibinfo{author}{Hadjiconstantinou,
  N.~G.}
\newblock \bibinfo{title}{An alternative approach to efficient simulation of
  micro/nanoscale phonon transport}.
\newblock \emph{\bibinfo{journal}{Appl. Phys. Lett.}}
  \textbf{\bibinfo{volume}{101}}, \bibinfo{pages}{153114}
  (\bibinfo{year}{2012}).

\bibitem{Peraud2014}
\bibinfo{author}{Peraud, J.-P.~M.}, \bibinfo{author}{Landon, C.~D.} \&
  \bibinfo{author}{Hadjiconstantinou, N.~G.}
\newblock \bibinfo{title}{Monte carlo methods for solving the boltzmann
  transport equation}.
\newblock \emph{\bibinfo{journal}{Annual Review of Heat Transfer}}
  \textbf{\bibinfo{volume}{17}} (\bibinfo{year}{2014}).

\bibitem{Peraud2014a}
\bibinfo{author}{Peraud, J.-P.~M.}, \bibinfo{author}{Landon, C.~D.} \&
  \bibinfo{author}{Hadjiconstantinou, N.~G.}
\newblock \bibinfo{title}{Deviational methods for small-scale phonon
  transport}.
\newblock \emph{\bibinfo{journal}{Mech. Eng. Rev.}}
  \textbf{\bibinfo{volume}{1}}, \bibinfo{pages}{FE0013--FE0013}
  (\bibinfo{year}{2014}).

\bibitem{Li2014}
\bibinfo{author}{Li, W.}, \bibinfo{author}{Carrete, J.~C.},
  \bibinfo{author}{Katcho, N.~A.} \& \bibinfo{author}{Mingo, N.}
\newblock \bibinfo{title}{Shengbte: A solver of the boltzmann transport
  equation for phonons}.
\newblock \emph{\bibinfo{journal}{Comput. Phys. Commun.}}
  \textbf{\bibinfo{volume}{185}}, \bibinfo{pages}{1747 -- 1758}
  (\bibinfo{year}{2014}).

\bibitem{shengbte}
\bibinfo{author}{Li, W.}, \bibinfo{author}{Carrete, J.},
  \bibinfo{author}{Katcho, N.~A.} \& \bibinfo{author}{Mingo, N.}
\newblock \bibinfo{title}{Shengbte}.
\newblock \bibinfo{note}{Www.shengbte.org, Date of access: 01/07/2014}.

\bibitem{phonopy}
\bibinfo{author}{Togo, A.}
\newblock \bibinfo{title}{Phonopy}.
\newblock \bibinfo{note}{Http://phonopy.sourceforge.net/, Date of access:
  01/07/2014}.

\bibitem{Kresse1993}
\bibinfo{author}{Kresse, G.} \& \bibinfo{author}{Hafner, J.}
\newblock \bibinfo{title}{Ab initio molecular dynamics for liquid metals}.
\newblock \emph{\bibinfo{journal}{Phys. Rev. B}} \textbf{\bibinfo{volume}{47}},
  \bibinfo{pages}{558--561} (\bibinfo{year}{1993}).

\bibitem{Kresse1994}
\bibinfo{author}{Kresse, G.} \& \bibinfo{author}{Hafner, J.}
\newblock \bibinfo{title}{Ab initio molecular-dynamics simulation of the
  liquid-metal–amorphous-semiconductor transition in germanium}.
\newblock \emph{\bibinfo{journal}{Phys. Rev. B}} \textbf{\bibinfo{volume}{49}},
  \bibinfo{pages}{14251--14269} (\bibinfo{year}{1994}).

\bibitem{Kresse1996}
\bibinfo{author}{Kresse, G.} \& \bibinfo{author}{Furthmüller, J.}
\newblock \bibinfo{title}{Efficiency of ab-initio total energy calculations for
  metals and semiconductors using a plane-wave basis set}.
\newblock \emph{\bibinfo{journal}{Comput. Mater. Sci.}}
  \textbf{\bibinfo{volume}{6}}, \bibinfo{pages}{15 -- 50}
  (\bibinfo{year}{1996}).

\bibitem{Kresse1996a}
\bibinfo{author}{Kresse, G.} \& \bibinfo{author}{Furthm\"uller, J.}
\newblock \bibinfo{title}{Efficient iterative schemes for ab initio
  total-energy calculations using a plane-wave basis set}.
\newblock \emph{\bibinfo{journal}{Phys. Rev. B}} \textbf{\bibinfo{volume}{54}},
  \bibinfo{pages}{11169--11186} (\bibinfo{year}{1996}).

\bibitem{Hua2014PRB2}
\bibinfo{author}{Hua, C.} \& \bibinfo{author}{Minnich, A.~J.}
\newblock \bibinfo{title}{Analytical green's function of the multidimensional
  frequency-dependent phonon boltzmann equation}.
\newblock \emph{\bibinfo{journal}{Phys. Rev. B}} \textbf{\bibinfo{volume}{90}},
  \bibinfo{pages}{214306} (\bibinfo{year}{2014}).

\bibitem{Wang2010}
\bibinfo{author}{Wang, Z.} \& \bibinfo{author}{Mingo, N.}
\newblock \bibinfo{title}{Diameter dependence of sige nanowire thermal
  conductivity}.
\newblock \emph{\bibinfo{journal}{Appl. Phys. Lett.}}
  \textbf{\bibinfo{volume}{97}}, \bibinfo{pages}{101903}
  (\bibinfo{year}{2010}).

\bibitem{Garg2011}
\bibinfo{author}{Garg, J.}, \bibinfo{author}{Bonini, N.},
  \bibinfo{author}{Kozinsky, B.} \& \bibinfo{author}{Marzari, N.}
\newblock \bibinfo{title}{Role of disorder and anharmonicity in the thermal
  conductivity of silicon-germanium alloys: A first-principles study}.
\newblock \emph{\bibinfo{journal}{Phys. Rev. Lett.}}
  \textbf{\bibinfo{volume}{106}}, \bibinfo{pages}{045901}
  (\bibinfo{year}{2011}).

\bibitem{Ziman1962}
\bibinfo{author}{Ziman, J.}
\newblock \emph{\bibinfo{title}{Electrons and Phonons: The Theory of Transport
  Phenomena in Solids}}.
\newblock International series of monographs on physics
  (\bibinfo{publisher}{Clarendon Press}, \bibinfo{year}{1962}).

\bibitem{Azeredo2013}
\bibinfo{author}{Azeredo, B.~P.} \emph{et~al.}
\newblock \bibinfo{title}{Silicon nanowires with controlled sidewall profile
  and roughness fabricated by thin-film dewetting and metal-assisted chemical
  etching}.
\newblock \emph{\bibinfo{journal}{Nanotechnology}}
  \textbf{\bibinfo{volume}{24}}, \bibinfo{pages}{225305}
  (\bibinfo{year}{2013}).

\bibitem{Feser2012}
\bibinfo{author}{Feser, J.~P.} \emph{et~al.}
\newblock \bibinfo{title}{Thermal conductivity of silicon nanowire arrays with
  controlled roughness}.
\newblock \emph{\bibinfo{journal}{J. Appl. Phys.}}
  \textbf{\bibinfo{volume}{112}}, \bibinfo{pages}{114306}
  (\bibinfo{year}{2012}).

\bibitem{Turney2010}
\bibinfo{author}{Turney, J.~E.}, \bibinfo{author}{McGaughey, A. J.~H.} \&
  \bibinfo{author}{Amon, C.~H.}
\newblock \bibinfo{title}{In-plane phonon transport in thin films}.
\newblock \emph{\bibinfo{journal}{J. Appl. Phys.}}
  \textbf{\bibinfo{volume}{107}}, \bibinfo{pages}{024317}
  (\bibinfo{year}{2010}).

\bibitem{Cai2010}
\bibinfo{author}{Cai, W.} \emph{et~al.}
\newblock \bibinfo{title}{Thermal transport in suspended and supported
  monolayer graphene grown by chemical vapor deposition}.
\newblock \emph{\bibinfo{journal}{Nano Lett.}} \textbf{\bibinfo{volume}{10}},
  \bibinfo{pages}{1645--1651} (\bibinfo{year}{2010}).

\bibitem{Chen2011}
\bibinfo{author}{Chen, S.} \emph{et~al.}
\newblock \bibinfo{title}{Raman measurements of thermal transport in suspended
  monolayer graphene of variable sizes in vacuum and gaseous environments}.
\newblock \emph{\bibinfo{journal}{ACS Nano}} \textbf{\bibinfo{volume}{5}},
  \bibinfo{pages}{321--328} (\bibinfo{year}{2011}).

\bibitem{Chen2012}
\bibinfo{author}{Chen, S.} \emph{et~al.}
\newblock \bibinfo{title}{Thermal conductivity of isotopically
  modified graphene}.
\newblock \emph{\bibinfo{journal}{Nat. Mater.}} \textbf{\bibinfo{volume}{11}},
  \bibinfo{pages}{203--207} (\bibinfo{year}{2012}).

\bibitem{Lindsay2010}
\bibinfo{author}{Lindsay, L.}, \bibinfo{author}{Broido, D.~A.} \&
  \bibinfo{author}{Mingo, N.}
\newblock \bibinfo{title}{Flexural phonons and thermal transport in graphene}.
\newblock \emph{\bibinfo{journal}{Phys. Rev. B}} \textbf{\bibinfo{volume}{82}},
  \bibinfo{pages}{115427} (\bibinfo{year}{2010}).

\end{thebibliography}

\begin{figure}[H]
\centering
  \includegraphics[width=0.99\columnwidth]{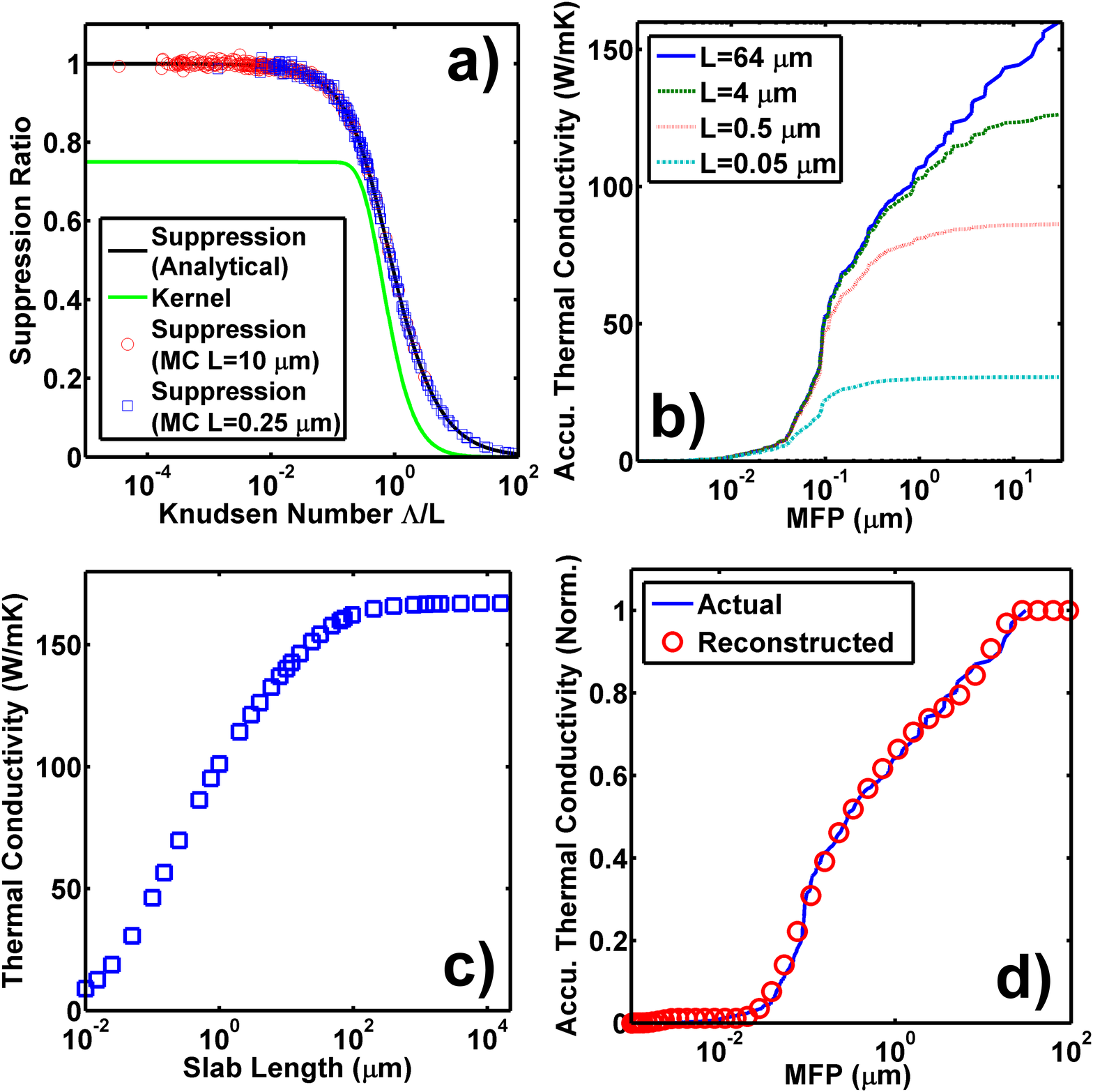}
  \caption{(a) Suppression functions obtained from analytical (black solid line) and MC methods (open circles and open squares), and the Kernel (green solid line)  from the analytic method. The two suppression functions are in excellent agreement.  (b) MFP spectra for pure Si slabs of various thicknesses obtained from MC. Larger suppression for long MFP phonons occurs as the length of the Si slab decreases.(c) Thermal conductivities of pure Si slabs as a function of thickness calculated by MC. (d) Reconstructed (red circles) and the actual (blue solid line) MFP spectrum of bulk Si. All the values are normalized to bulk thermal conductivity of pure Si. The reconstructed result is in excellent agreement with the actual MFP spectrum.}
\end{figure}

\begin{figure}[H]
\centering
  \includegraphics[width=0.99\columnwidth]{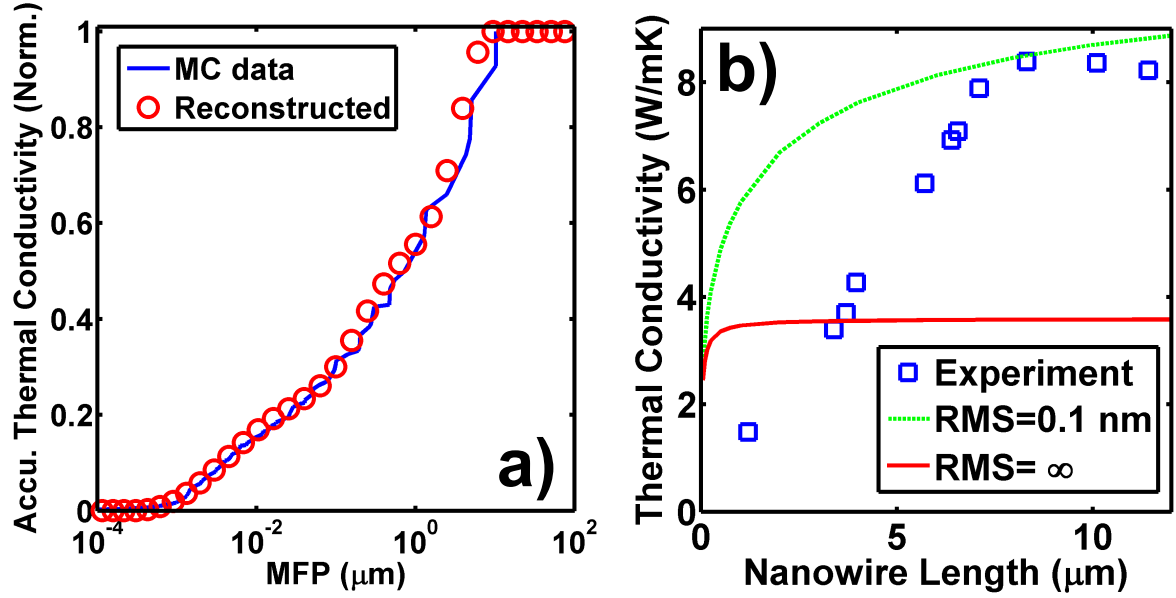}
  \caption{(a)  The actual (MC data) and reconstructed MFP spectra of a simulated Si$_{0.9}$Ge$_{0.1}$ nanowire with surface roughness of RMS = 0.1 nm and the same square cross-section with a side length of 100 nm. The reconstructed distribution is in good agreement with the actual distribution, even though it is extracted from merely a series of discrete thermal conductivities without any boundary scattering information. (b)Thermal conductivities of Si$_{0.9}$Ge$_{0.1}$ nanowire from both MC simulations (lines) and experimental measurements  \cite{Hsiao2013} (open squares) as a function of nanowire lengths.  All simulated nanowires have the same square cross section as that in (a), and were simulated with very smooth surface ($\sigma$ = 0.1 nm, green dashed line) and extremely rough surface ($\sigma$ = infinity, red solid line), respectively. The experimental data does not follow the trend predicted by the simulations. All the thermal conductivities are normalized to their ``bulk" value, which is thermal conductivity of the infinitely long nanowire.}
\end{figure}

\begin{figure}[H]
\centering
  \includegraphics[width=\columnwidth]{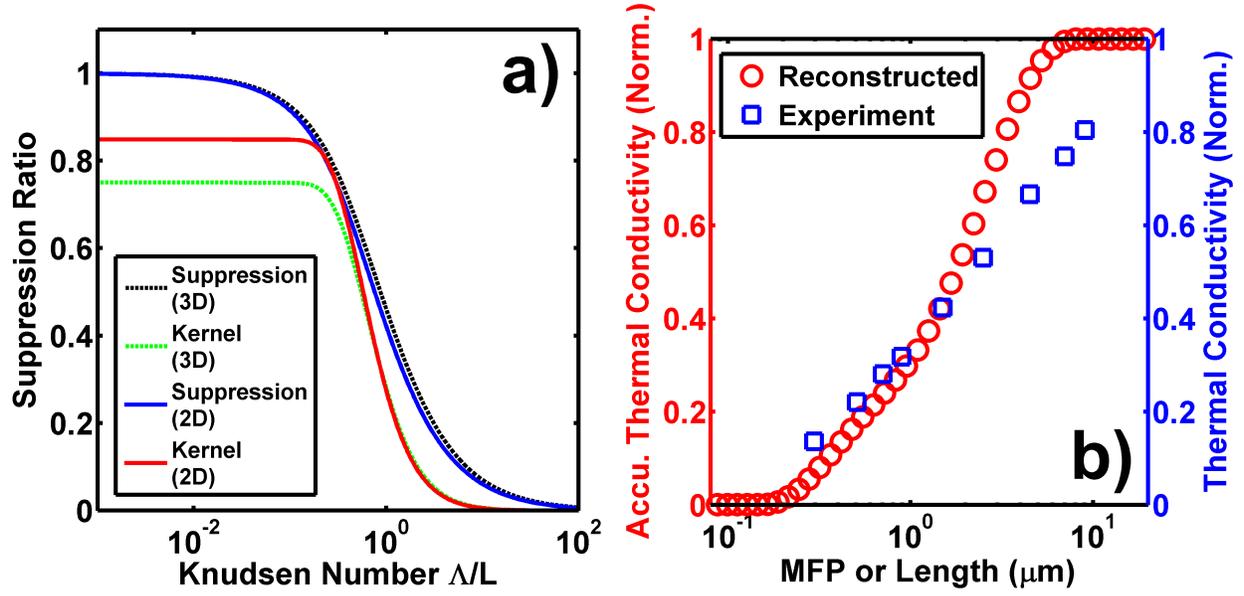}
  \caption{(a) Suppression function in both 3D (dashed black line) and 2D (solid blue line) spaces and their corresponding kernel functions. The 2D suppression function is very similar to the 3D one. (b) Experimentally measured length dependent thermal conductivity\cite{Xu2014} (blue open squares) and the corresponding reconstructed accumulative thermal conductivity as a function of phonon MFP (red open circles) in suspended graphene samples. All these thermal conductivities are normalized to thermal conductivities of ``bulk" graphene flakes, which is calculated using the extrapolating method in Ref. \citenum{Wei2014}. Phonons with MFPs longer than 1 $\mu$m carry the majority of the heat in suspended graphene.}
\end{figure}

\end{document}